\documentclass[12pt]{article}

\def\be{\begin{eqnarray}}
\def\ee{\end{eqnarray}}
\def\bq{\begin{equation}}
\def\eq{\end{equation}}
\def\ben{\begin{enumerate}}\def\een{\end{enumerate}}

\def\roughly#1{\mathrel{\raise.3ex\hbox{$#1$\kern-.75em%
\lower1ex\hbox{$\sim$}}}}

\usepackage{epsfig}

\begin{document}
\begin{titlepage}

\hfill FTUV-05-0929

 \vspace{1.5cm}
\begin{center}
\ \\
{\bf\LARGE Bursts of pions as a signature of Quark Gluon Plasma}
\\
\vspace{0.7cm} {\bf\large Vicente Vento} \vskip 0.7cm

{\it  Departamento de F\'{\i}sica Te\'orica and Instituto de
F\'{\i}sica Corpuscular}

{\it Universidad de Valencia - Consejo Superior de Investigaciones
Cient\'{\i}ficas}

{\it 46100 Burjassot (Val\`encia), Spain, }

{\small Email: Vicente.Vento@uv.es}

\end{center}
\vskip 1cm \centerline{\bf Abstract} \vskip 0.3cm

We study the behavior of the scalar glueball inside a Quark Gluon
Plasma. We follow the fireball from the plasma phase to the
hadronic phase and observe that an interesting phenomenon, bursts
of pions describing the flow of matter, occurs which might have
experimental consequences as a signature of Quark Gluon Plasma.

 \vspace{2cm}

\noindent Pacs: 14.80.-j, 24.80.+y, 25.75.Nq

\noindent Keywords:  quark, gluon, glueball, meson, plasma

\end{titlepage}


\indent\indent Quantum Chromodynamics (QCD)  is the theory of the
strong interactions \cite{FritzschGellMannLeutwyler}. At low
temperatures its elementary constituents are mesons and glueballs
\cite{MinkowskiFritzsch}, this is the hadronic phase where all
states have no color, i.e. they are color singlets. On the
contrary, at high temperatures a phase transition, called
de-confinement (liberation of color), takes place and its
elementary constituents become color quarks and gluons. This phase
is known as the Quark Gluon Plasma (QGP)\cite{QGP}. Our aim is to
study the behavior of QCD in the transition from the QGP to the
hadronic phase centering our attention in the behavior of the
scalar glueballs. These particles are bound states of gluons, the
gauge bosons of QCD. This unique structure led to an intense
experimental search, since they were first theoretically
contemplated \cite{MinkowskiFritzsch}, which has not produced yet
a clear picture of their spectrum . Recently, we have proposed an
interpretation of the scalar particles that contemplates a rich
low lying glueball spectrum \cite{Vento}. We describe here a
scenario for the transition from Quark Gluon Plasma (QGP)  to
hadronic matter which arises from the behavior of the lightest
glueball, the scalar $0^{++}$, as the fireball cools  in heavy ion
collisions \cite{RHIC}. Our scenario manifests itself
experimentally as peculiar bursts of pions describing the matter
flow in the fluid created by the two colliding ions.

The realization of scale symmetry in Gluodynamcis (GD), the theory
with gluons and no quarks, provides a relation between the
parameters of the lightest scalar glueball, hereafter called $g$,
and the gluon condensate \cite{Schechter,Migdal,EllisLanik},

\bq m_g^2 f_g^2 = - 4 \; <0| \; \frac{\beta(\alpha_s)}{4
\alpha_s}\; G^2\;|0>,\label{mgfg} \eq
where $f_g = <0|g|0>$ , $m_g$ the $g$ mass, and the right hand
side arises from the scale anomaly. GD provides a description for
glueballs which almost coincides with that of QCD in the limit
when the OZI rule is exactly obeyed, i.e., when decays into quarks
which require gluons are strictly forbidden. This limit we have
called OZI Dynamics (OZID) \cite{Vento}.

Lattice results \cite{Lattice,Langfeld} and model calculations
\cite{Heinz,Drago,Wambach} support the traditional scenario
 \cite{Leutwyler,Miller}, that the condensate is basically
constant up to the phase transition temperature $T_C$ ($150 MeV <
T_C < 300 MeV$) and decreases slowly thereafter until it dilutes
(or evaporates) into gluons at $(2-3)  T_C $. In this regime the
mass of $g$ changes slowly across the phase transition
\cite{Heinz, Drago,Wambach} and might even increase beyond $T_C$
as the gluon binding energy decreases \cite{Shuryak}. These
results and Eq.(\ref{mgfg}) determine that $f_g$ will be small
only close to the dilution temperature when, in GD, scale
invariance is restored. However, around $T_C$, $f_g$ is sizeable
and therefore we are able to use in the scalar sector the OZID
approximation of QCD, where glueballs and mesons are almost
decoupled, and therefore the scalar glueballs of QCD behave
similarly to those of GD \cite{Vento}.

We assume for our discussion a recent formulation of the dynamics
in the region above $T_C$, which states that despite
de-confinement the color Coulomb interaction between the
constituents is strong and a large number of binary (even color)
bound states, with a specific mass pattern, are formed
\cite{Shuryak}. With this input, the scenario we envisage for GD
goes as follows. The strong Coulomb phase is crowded with gluon
bound states and $g$ is the lightest. As we move towards the
dilution limit, the binding energy of these states decreases, the
gluon mass increases, and therefore the color and singlet bound
states increase their mass softly until the gluons are liberated
forming a liquid \cite{Shuryak,Polyakov,Linde}. However, as we
cool towards the confining phase, color and singlet states decay
into the conventional low lying glueballs, in particular $g$. The
fact which makes this scenario appealing is that the multiplicity
of glueball channels grows tremendously above the phase
transition. The ratio of glueball to meson channels goes from 1 to
8 below the phase transition to 1 to 2 above. We expect the number
of glueballs to be very large in the cooling of the fireball. We
are not the first to single out glueballs as a possible signature
of QGP \cite{Shuryak0} but certainly both mechanisms and their
consequences are completely diverse.

When two heavy ions collide at ultra-relativistic energies, if the
collision is quite central, a hot region of space time is produced
called the fireball \cite{RHIC}. Let us incorporate in the cooling
of the fireball the dynamics of QCD as described in our scenario.
Our starting point is a plasma with a temperature $T_C < T <
3T_C$. This plasma is almost a perfect fluid of hadronic matter
with low viscosity and is full of binary states \cite{Shuryak}.
The lowest mass $q \bar{q}$ states are the pseudoscalar pion $\pi$
and the scalar meson $\sigma$, which are here bound states of the
strong color interaction. The lightest glueball state is $g$. The
behavior of $g$ runs together with all other hadronic processes
leading to a collective flow but, in the OZID approximation, it
can be singled out.

As the fireball cools the large number of gluonic bound states
decay by gluon emission into $g$'s. The emitted gluons form new
bound states of lower mass, due to the strong color Coulomb
interaction. As we approach the confinement region the mass of the
color bound states increases and it pays off to make multiparticle
color singlet states, which decay by rearrangement into ordinary
color singlet states. Since the coupling is strong and the phase
space is large, these processes take place rapidly. Thus in no
time, close to the phase transition temperature $T_C$, a large
number of scalar glueballs populate the hadronic liquid. In our
idealized OZID world they interact among themselves and with quark
matter only by multi gluon exchanges, i.e., weak long range color
Van der Waals forces. Color factors make the $g-g$ interaction
stronger than the $g-q$  $(g-\bar{q})$ one. Thus the former
produce droplets of a glueball liquid within a background of
hadronic liquid and the weak residual interaction between the
glueballs and quark matter makes these $g$-droplets be dragged by
the hadronic liquid with the flow determined by the kinetics of
the binary states from which they all proceed.

 The $g$-droplets have a large mean free path
since they interact weakly with quark matter and therefore as the
liquid slows down by the increase of the hadronic interactions the
$g$-droplets escape from the liquid flowing transversally faster,
at which point they start to encounter other droplets and
percolate into larger and larger droplets. Thus we arrive to a
geometry of $g$-droplets following the flow of hadronic matter.

The scalar glueball $g$ decays into pions by mixing with a scalar
$\sigma$ meson \cite{Vento}. If we assume that the $\sigma$ is the
O(4) partner of the $\pi$ in the chiral symmetric realization of
QCD, its mass starts to decrease before the phase transition,
becoming degenerate with the pion . Moreover, since in the strong
Coulomb phase chiral symmetry is restored, it remains degenerate
thereafter \cite{Shuryak} (see Fig.1). The mass of $g$ does not
vary in this region appreciably. Thus even before we reach $T_C$
the mixing between $g$ and $\sigma$ disappears because their
masses become very different (see Fig.1) . Therefore $g$ becomes
stable. This stability remains in the strong color Coulomb phase,
since when the mass of the $\sigma$ increases to the point of
allowing mixing, its degeneracy with the pion does not allow for
the $2\pi$ decay process (see Fig.1) .

\vskip 0.5cm
\begin{figure}[htb]
\centerline{\epsfig{file=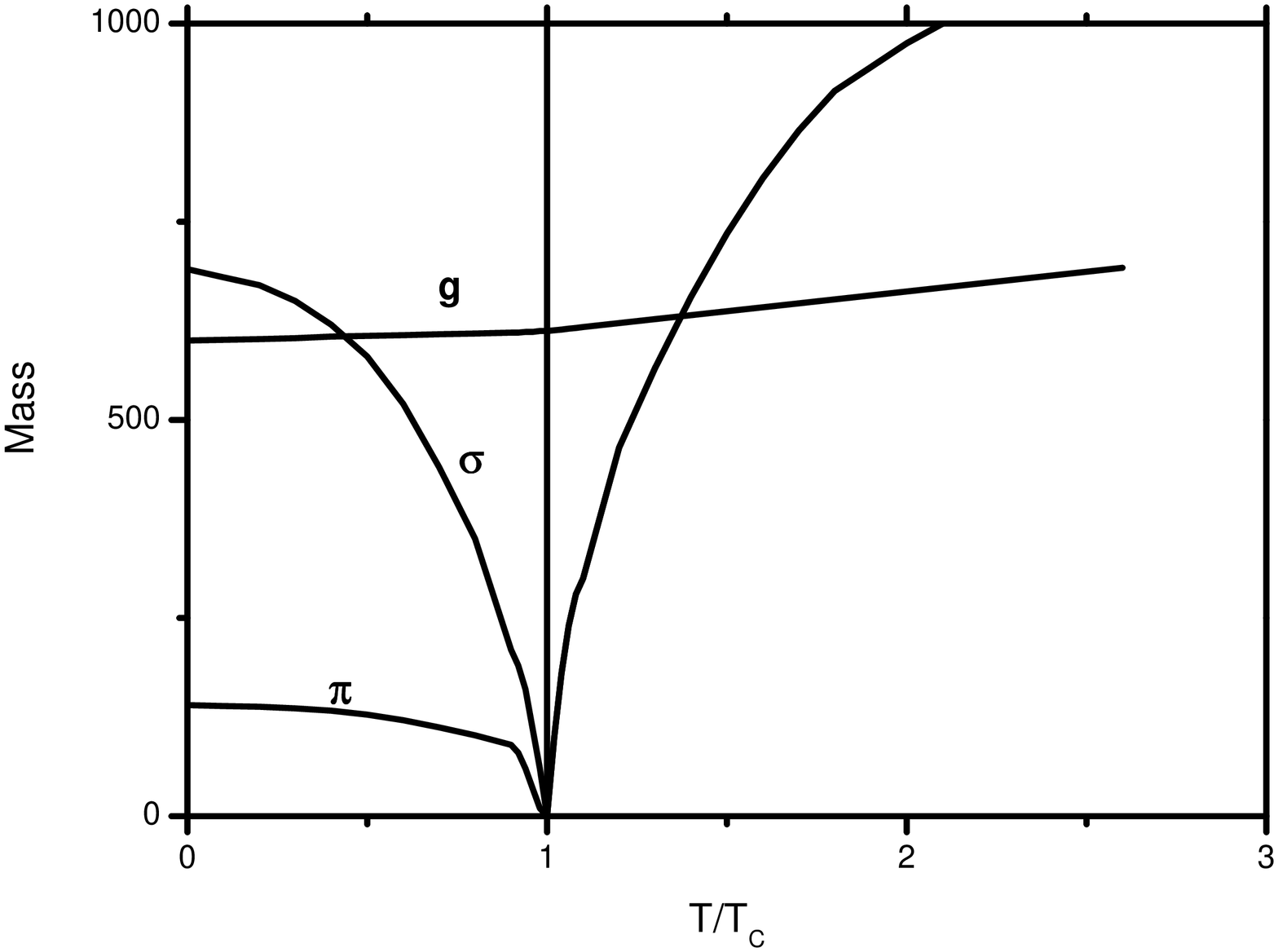,width=7.0cm,angle=270}}
\caption{Schematic representation of the behavior of the masses of
$\sigma$,$\pi$ and $g$ across the QGP phase transition.} \vskip
0.5cm
\end{figure}

Now we have all the ingredients to discuss the observational
signatures of our scenario. From the previous discussion it is
clear that the $g$-droplets are stable in the strong color Coulomb
phase. As the fireball cools, the $g$'s cross the phase transition
point as stable states and only when the value of the $\sigma$
mass increases close to  that of the $g$ ($T < T_C/2$), certainly
higher than $2m_\pi$, decay may take place (see Fig.1). This
mechanism provides us with a ``long time span" after the phase
transition. Thereafter mixing occurs, however in the approximation
of ref.{\cite{Vento}, $\Gamma_g << \Gamma_\sigma \sim 100$ MeV,
since the latter is a typical hadronic width. Thus the
$g$-droplets ``live much longer" than the conventional hadronic
states and therefore they are the last states to decay. Moreover,
since the $g$-states have been formed in a determined geometry,
namely droplets following the elliptic flow, we expect almost
simultaneous bursts of pions arising from the droplets covering
the topology of the detector and describing the behavior of the
flow. If the $g$-droplets condense coherence would lead to more
spectacular bursts.

Thus our observational signature of glueballs in QGP is a
distribution of pionic bursts following the ellipsoidal flow
arising from droplets, which happen after all other pionic
emissions have taken place. Since the droplets, up to the point of
decay, have been flowing faster than the rest of the hadronic
medium the bursts maintain the well defined production geometry.

How to detect this signature does not look easy. If we could make
a ``movie", by binning the time periods of arrival of pions, the
lasts bins would contain the proposed bursts. Moreover,  since
half of the pions that will come out of the decay are $\pi^0$'s
one could attempt to correlate the pionic bursts with gamma bursts
arising from $\pi^0 \rightarrow 2\gamma$.

Let us conclude by stating that we have analyzed the behavior of a
peculiar hadronic state, the scalar glueball, in a hot hadronic
medium. We have discussed in physical terms how these particles,
which are created copiously in the strong color Coulomb regime,
behave as the fireball cools down. We have seen that their weak
coupling with other hadronic states provide them with a well
defined behavior in the plasma as the temperature drops. This
behavior is transferred to its detectable decay products leading
to a peculiar emission of pions (and photons) in the forms of
bursts of particles, which hints a possible signature for QGP
formation. Our most important result is, that the stability of the
glueball across the phase transition up to the point where mixing
with the sigma occurs, together with the subsequent small decay
width into pions confer to the realization of our proposed
observation plausibility.

\subsection*{Acknowledgments}

I would like to thank K. Langfeld and B.-J. Schaefer for
clarifying information regarding their work. I acknowledge the
hospitality extended to me by the GSI theory group, were this
research was initiated, and in particular the conversations with
Bengt Friman were illuminating.

\end{document}